# A Novel AI-enabled Framework to Diagnose Coronavirus COVID-19 using Smartphone Embedded Sensors: Design Study

Halgurd S. Maghdid, Kayhan Zrar Ghafoor, Ali Safaa Sadiq, Kevin Curran, Danda B. Rawat, Khaled Rabie

*Abstract*—Coronaviruses are a famous family of viruses that cause illness in both humans and animals. The new type of coronavirus COVID-19 was firstly discovered in Wuhan, China. However, recently, the virus has widely spread in most of the world and causing a pandemic according to the World Health Organization (WHO). Further, nowadays, all the world countries are striving to control the COVID-19. There are many mechanisms to detect coronavirus including clinical analysis of chest CT scan images and blood test results. The confirmed COVID-19 patient manifests as fever, tiredness, and dry cough. Particularly, several techniques can be used to detect the initial results of the virus such as medical detection Kits. However, such devices are incurring huge cost, taking time to install them and use. Therefore, in this paper, a new framework is proposed to detect COVID-19 using built-in smartphone sensors. The proposal provides a low-cost solution, since most of radiologists have already held smartphones for different daily-purposes. Not only that but also ordinary people can use the framework on their smartphones for the virus detection purposes. Today's smartphones are powerful with existing computation-rich processors, memory space, and large number of sensors including cameras, microphone, temperature sensor, inertial sensors, proximity, colour-sensor, humidity-sensor, and wireless chipsets/sensors. The designed Artificial Intelligence (AI) enabled framework reads the smartphone sensors' signal measurements to predict the grade of severity of the pneumonia as well as predicting the result of the disease.

## I. Introduction

A novel coronavirus infection disease named COVID-19 was first identified in Wuhan city, Hubei Province of China, China. In December 2019, World Health Organization (WHO) announced that the virus can cause respiratory disease with manifesting cough, fever and pneumonia. The disease has spread in China and has now been identified in many countries internationally [1], [2]. The Emergency committee of WHO on January 30, 2020 declared this virus spread as a pandemic disease because of its fast person-to-person spread and the fact that most of infected people are not immune to it.

The COVID-19 is common in people and many different species of animals, including camels, cattle, cats, and bats. Firstly, infected people with the novel COVID-19 at the epicenter in Wuhan had links with seafood and live animal markets, indicating animal-to-person spread. Thereafter, rising numbers of patients who did not have contact with the live animals, resulted in person-to-person spread [3]. Subsequently, the WHO on March 11, 2020 announced the novel COVID-19 outbreak a pandemic as on that date the number of confirmed cases reached 118,000 with more than 4000 deaths [4].

The COVID-19 and human coronaviruses are categorised under the family of Coronaviridae. These viruses infect people with moderate cold Middle East Respiratory Syndrome (MERS) or Severe Acute Respiratory Syndrome (SARS) [5]. SARS is also a viral respiratory disease caused by SARS- associated coronavirus (SARS-CoV), which was first reported in 2003 in Southern China and spread in many countries worldwide. Moreover, MERS virus cases were first announced in Saudi Arabia and cause the death of 858. Based on the analysis of virus genomes, this virus was originated in bats [6]. The clinical presentation of the COVID-19 is complicated and could be manifested as fever, cough and severe headache. There are several techniques for COVID-19 detection, such as the Nucleic Acid Test (NAT) and Computed Tomography (CT) scan. The NAT is utilized to detect specific nucleic acid sequence and species of organism, predominately a virus or bacteria that can cause disease in blood, tissue or urine. Although NAT technique and detection kits are significant for COVID-19 virus detection, CT scan is the most effective and functional for detecting the severity and degree of the lung inflammation [7]. The National Health Commission of China confirmed the inclusion of radiographic presentation of pneumonia for clinical diagnostic standard in Hubei Province [8].
This assures the significance of the CT scan images for the diagnosis of COVID-19 pneumonia severity.

Recently, WHO reported COVID-19 as a pandemic and thousands of patients have been spending hours waiting at hospitals for CT scan image examination. This does not only overload the medical system and get the patients frustrated,

Halgurd S. Maghded is with the Department of Software Engineering, Faculty of Engineering, Koya University, Kurdistan Region-F.R.Iraq. First.Last@koyauniversity.org.

Kayhan Zrar Ghafoor is with the Department of Software Engineering, Salahaddin University-Erbil, Iraq; School of Mathematics and Computer Science, University of Wolverhampton, Wulfruna Street, Wolverhampton, WV1 1LY, UK. kayhan@ieee.org.

Ali Safaa Sadiq is with the School of Mathematics and Computer Science, University of Wolverhampton, Wulfruna Street, Wolverhampton, WV1 1LY, UK. Centre for Artificial Intelligence Research and Optimisation, Torrens University Australia, Fortitude Valley, Brisbane, 4006 QLD, Australia. Ali.Sadiq@wlv.ac.uk.

Kevin Curran is with the School of Computing, Eng & Intel. Sys, Ulster University, Londonderry, BT48 7JL, UK. kj.curran@ulster.ac.uk.

Danda B. Rawat is with the Department of Electrical Engineering and Computer Science at Howard University, Washington, DC 20059, USA. db.rawat@ieee.org

Khaled Rabie is with the Department of Electrical and Electronic Engineering, Manchester Metropolitan University, UK. K.Rabie@mmu.ac.uk

but also leads to a serious risk of cross-infection with other patients. Particularly, at the province of Hubei, suspected cases, confirmed COVID-19 patients and cases under medical observation need to go through CT imaging of lung. Besides, the number of radiologists is comparably much less than the number of patients. As a result, this makes medical systems and physicians overloaded. This leads to late detection and quarantine of infected people and less efficient treatment of patients [8]. More precisely, recently in Italy, hospitals are only received high priority people, who have huge fever and shortness of breathing [9].

The pandemic of COVID-19 and the resultant huge demand for diagnosis has driven companies, academics and researchers to provide high responsive, intelligent and more efficient detection methods. Ping a Smart Healthcare company, revealed an intelligent technique for COVID-19 smart CT image reading that can analyse results in about 15 seconds with an accuracy rate above 90% [10]. However, both Reverse Transcription Polymerase Chain Reaction (RT-PCR) and CT scan for COVID-19 diagnosis are not perfect [10]. Thus, the most reliable technique is the combination of several methods. Furthermore, medical detection kits are used to detect the COVID-19, but such devices are costly and require installation for diagnosis.

Modern smartphones are embedded with numerous sensors with powerful computation capabilities. Using smartphones, it is possible to sense information about daily activities and even capturing visual data [11]. One of the important features of smartphones is the capability of capturing, collecting and storing large volumes of data from either suspected and confirmed COVID-19 cases. In particular, a smartphone has a capability to scan CT images of a COVID-19 patients for analysis purposes. Further, multiple CT images of the same COVID-19 patient can be fed to the smartphone for comparative analysis on how lesions have been developed. The analysis is very useful to the suspected COVID-19 cases to diagnosis and monitor the grade of lung inflammation.

In this paper, we present for the first time a new framework to diagnose the COVID-19 using the built-in smartphone sensors. The framework provides a low cost solution, since most of the radiologists have smartphones used for different daily-purposes. Not only that, but also ordinary people can use the framework on their smartphones for COVID-19 diagnosis. Moreover, people can use the proposed smartphone-based framework to monitor their grade of lung inflammation. Today's smartphones are embedded with computation-rich processors, memory space, and many sensors including cameras, microphone, temperature sensor, inertial sensors, proximity, color-sensor, humidity-sensor, and wireless chipsets/sensors. The designed framework allows to read the smartphone sensors' signal measurements and scans CT images to identify viral pneumonia. The developed framework takes much less time to identify COVID-19 as compared to the expert radiologist. The radiologists also can use the framework to track the development of the disease grades and evaluate for treatment. The rest of this paper is organized as follows. Section

II explains the background of the literature on developed AI systems for COVID-19 detection. This is followed by presenting an overview of the proposed approach and details of the designed algorithm. Finally, Section IV concludes the paper.

## II. BACKGROUND

This section presents a review of very recent existing literature on the techniques used for novel COVID-19 diagnosis. There are several ways to identify the viral pneumonia on suspected cases.

Although very scarce literature exists on the diagnosis COVID-19, because of its new emergence, there are a few state-of-the-art reviews in this field. In an attempt, the authors in [1] developed an AI engine based on deep learning to detect COVID-19 using high resolution CT images. However, their proposed model only relies on CT images. Based on the latest research [10], the COVID-19 detection results are more reliable when several methods are used jointly. In another attempt, Ping An Insurance Company of China Ltd [10], developed a smart CT image reading system, which can read and diagnose in very short period of time.

## III. THE PROPOSED FRAMEWORK

This section presents the motivation and the detailed design of the proposed approach. The proposed approach elaborates the process of creating a smartphone-based framework, which includes smartphone, algorithms and embedded sensors.

The proposed framework and its associated algorithms utilise smartphone sensors to diagnose the preliminary results of the COVID-19 diagnosis. Although, there are several methods to get the result of the disease test, the proposal provides a low cost and friendly solution. The solution could be used by radiologists or people who have smartphones at anytime, anywhere. Therefore, such framework is needed useful in the emergency situations.

To understand the workflow of the framework, first, the symptoms of the confirmed COVID-19 patient should be realized. The well-known symptoms of the disease are: fever, fatigue, headache, nausea, dry-cough, lung CT imaging features, and shortness of breath. Each of these symptoms has its own level which could be differentiated from other disease including flue-symptoms, cold-symptoms and hay fever symptoms. For this reason, the framework tries to discover the level of each symptoms based on the built-in sensors measurements. In addition, a set of sensors technologies are embedded on the smartphones including cameras, inertial sensors, microphone, and temperature sensor. In addition, the readings of these sensors technologies have been used for those symptoms which exist in the coronavirus diseases. As state of the art, algorithms are separately applied on each of these sensors' readings to detect the level of the symptoms for the human-health purposes. For example, temperature-fingerprint sensor as it is located under the smartphone's touch-screen has been used to predict the fever level in [12]. Camera captured images and videos are used to detect human fatigue in different environment via human-gait analysis in [13]. Further, onboard inertial sensors (such as accelerometer sensors) have been used in [12] and [14] to detect the fatigue level. In [15], the nausea prediction has been analyzed based on smartphones-enabled video observation and directly observed treatments. In another vain, both the camera sensor and inertial sensors' measurements have been utilized in [16] to monitor neck posture and consequently to predict the level of human-headache. A comprehensive work has been studied to indicate the type of cough by using smartphone-microphone chipset in [17] and [18].

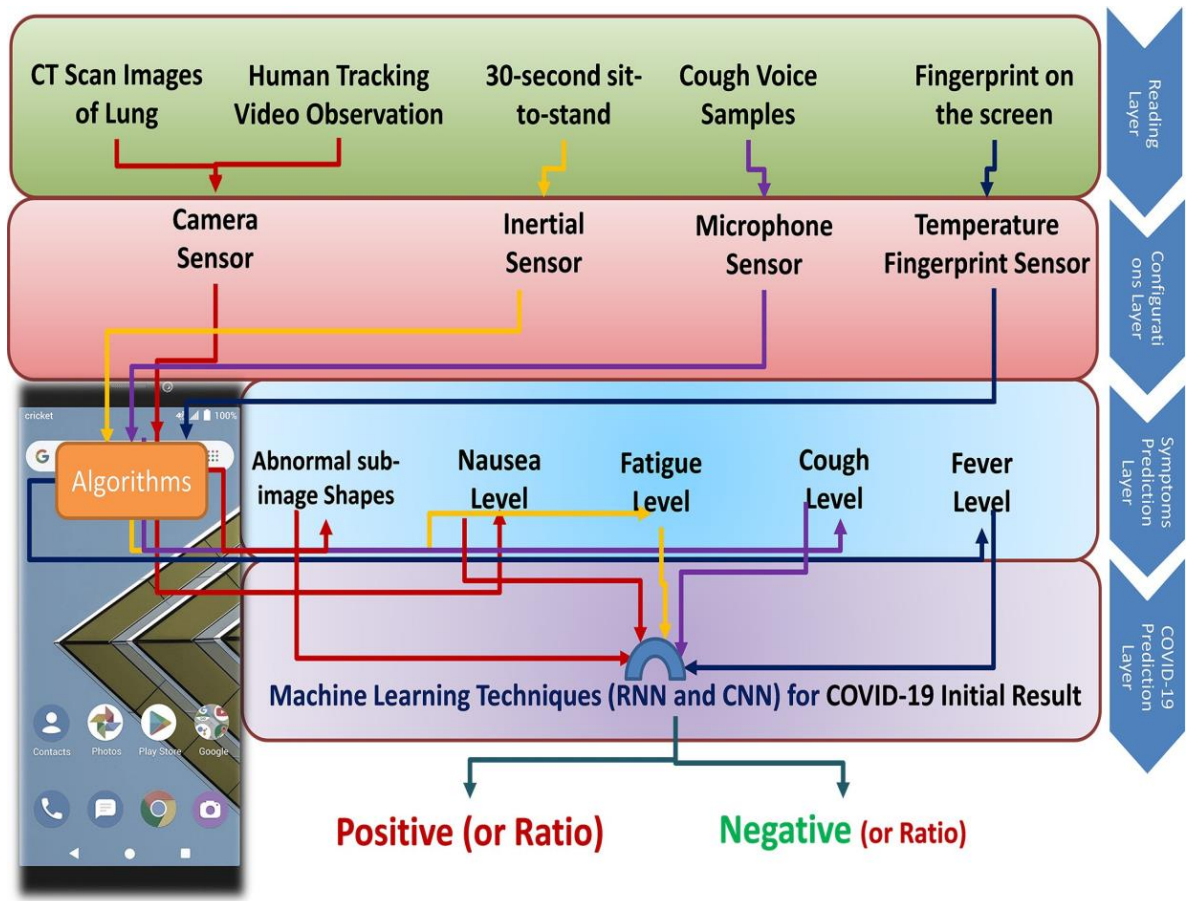

Fig. 1. General diagram of the proposed framework for predicting disease COVID-19.

In this article, the proposed framework aims to use all these aforementioned sensors and algorithms (could be improved) in a single solution. This is followed by getting the predicted level results of symptoms from the applied algorithms and stored them in a dataset as a single record. Thus, such records from different patients could be collected and used as input to a Machine Learning (ML) algorithm. There are several ML techniques used for human-health purposes such as decision tree, support vector machine, k-nearest neighbors and neural networks. The most recent and accurate techniques is deep learning scheme, also referred to as neural network family. Many deep learning algorithms have been utilized for the classification or recognition purposes such as convolution neural network (CNN) and recurrent neural network (RNN) algorithms. CNN is feed-forward neural network which is generally used for spatial data such as image recognition [19]. While with RNN, the output of each layer will be saved and then used as an input for the next layers as well as the RNN is efficient for temporal data such as text [20] and signal measurements [21].

The proposed framework consists as a set of layers, as shown in Fig. 1. The first layer functionalities are responsible for reading the data from the sensors. For example, reading the captured CT scan images of lung and videos through using the smartphone camera, getting the inertial sensors (accelerometer sensor) measurements during 30-second sit-to-stand, recording microphone voice measurements during a series of cough, and finally scanning temperature sensor measurements during fingerprint touching on the smartphone screen. The second layer is structured to configure the onboard smartphone sensors including reading intervals, image size, buffers' size, timer resolution, and etc. Further, the readings and configurations are then used as the input of the symptoms algorithms running on the smartphone application. The third layer of the framework provides a the calculated symptoms level, separately, and then stored as a record input to the next layer. The last layer is to apply ML techniques to predict the COVID-19. The ML techniques could be used according to the nature of the recorded data. For example, for the abnormal sub-image of the CT scan images, CNN could be used. This is followed by using RNN for the rest of the recorded data. Therefore, a new combined CNN and RNN ML technique can be proposed for the framework.

In the proposed framework, one significant input to the smartphone is the CT scan images of the pneumonia-probable cases. Certainly, CT scan images is a key method to detect COVID-19. An algorithm can be developed to diagnose the lesions of patients caused by COVID-19 and analyze its size and density. The algorithm can compare multiple CT images of the lung lesions. The most important output of the algorithm is the volume and density of lesions. The most compelling evidence of the confirmed COVID-19 case is an increase of density and volume of lesions in CT images. Certainly, comparing multiple CT images takes long time and its interpretation, by the radiologists, cannot accurately be concluded manually. Thus, the proposed framework assists radiologists and enables people to make efficient and confident decisions on the suspected cases.

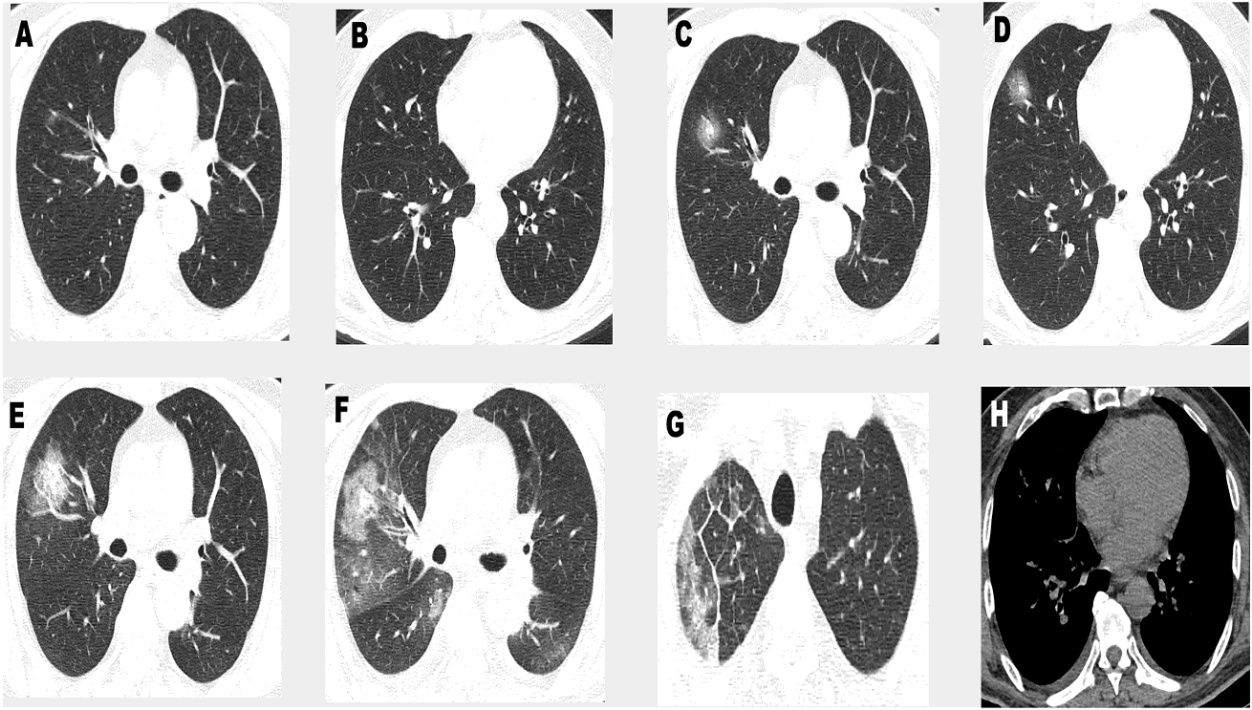

Fig. 2. CT images of the patient: opacities can be seen in the right lung of images A and B. 2 days later, an increase of opacities can be observed in images C and D. In sixth day, lesions can be observed. In images F, G, H, in ninth day, progressive lesions is shown in left and right lungs. Image F shows the appearance of small bilateral pleural effusions. [22].

In Fig. 2, the progressive effect of COVID-19 of a 61- year-old man admitted to the hospital in Lanzhou, China on January 25, 2020 due to close contact with a confirmed COVID-19 patient 10 days prior. The authors in [22], based on multiple CT images, described features of lung inflammation in a laboratory-confirmed COVID-19 patient. As can be seen, the volume quantity, grade and density of lesions and opacities are growing progressively from CT images A to G. Analysis of Fig. 2 by radiologists is very time consuming. In the proposed framework, the CT images of Fig. 2 is scanned and an algorithm can compute the volume and density of lesions and opacities in order to identify the stage of pneumonia in the course of the disease.

Further, to improve the proposed framework or to get a reliable prediction result, the recorded data and the result of the prediction could be exchanged in the cloud, as shown in Fig. 3. Thus, by using the framework from different users or patients, the framework dataset will grow and construct a large data set. Further, such process will provide transfer learning from multiple smartphone and various onboard sensors to the new smartphones.

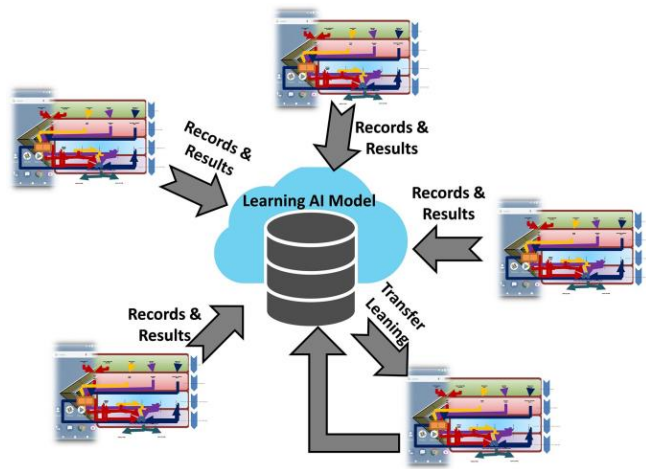

Fig. 3. Cloud computing for the proposed framework.

## IV. TITLE

This section presents the implemented COVID-19 diagnosing system. As proof-of-concept, a mobile APP based system was implemented to test the capturing functions of COVID-19 symptoms that used in diagnosing process, which are detailed in the previous section. Fig. 4 presents the screenshots of the registration page and submitting the captured mobile sensory data to the cloud platform.

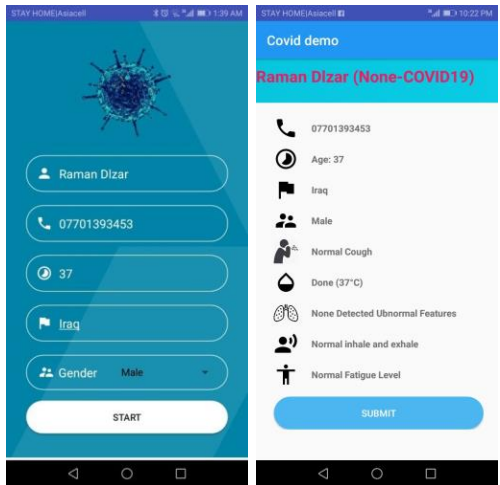

(a) User registration in the system  (b) The result of COVID-19 test

Fig. 4. User registration & results of the test

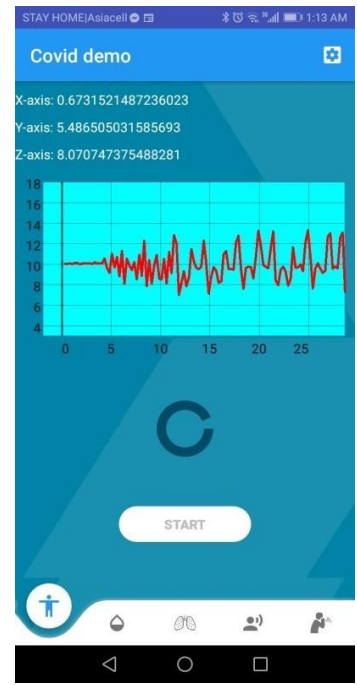

Fig. 5. Fatigue measurement.

In order to measure the fatigue level, an accelerometer sensory data measurements' function was implemented and the reading of these sensors are captured by this function as presented in Fig. 5. When a user activating this function, x, y and z-axes are captured to measure the stability of that user. This function can be activated at the background so that a time serious data will be obtained, which helps in making better decision on the fatigue level and construct an accurate pattern out of it

On the other hand, body temperature as stated in the previous section is considered to be one of the essential data inputs in diagnosing COVID-19; hence it was implemented and tested as shown by Fig. 6. Smartphone's rear camera sensor is utilized to calculate the body temperature based on placing the user's index finger on it so that our developed algorithm captures the body's temperature instantly. As it can be seen from Fig. 6, we have included four steps that showing a simple tutorial to the user of how making an accurate measure: i) notation on how the camera is used to get the body temperature, ii) the next step is to put the index finger on the back flash, iii) followed by the next step is the process of reading and converting the image features to heart rate and consequently to the temperature degree.

Fig. 7 demonstrates the implementation of X-ray/CT-Scan function that used in our proposed framework in diagnosing the COVID-19 pneumonia level. In our implemented method the scan, capture, and read functions of the X-ray or CT-scan images that captured of a patient's lung; are captured via smartphone's camera then stored on a cloud storage to be segmented and analyzed accordingly. There are two options to read the images of the X-ray and CT-scan, either directly via camera capture or by selecting the images that stored on the smartphone. The developed App is also providing the image process options to refine and augment images including resizing, cropping, converting to gray scale, converting to binary images, as well as making magic levels on the images. Another important input that would help in obtaining bet- ter diagnosing of COVID-19, breathing behavior analysis is implemented and captured in our developed App. Fig. 8 illustrates the a demo of breathing measure function, which shows the calculation of the breathing inhale and exhale durations. User need to capture multiple samples of inhale and exhale then obtain the average time take in each, which will be used in indicating the likelihood of COVID-19 infection. The microphone sensor is used also to capture the voice of breathing. Our developed system allows the medical doctor to set a timer to calculate the duration of the breathing process by a patient to be used in continuously monitoring the case. The process could be repeated for three times and then the average value of the duration for each inhale and exhale breathing will be taken.

Finally, the cough sound samples are captured by our developed function of our smartphone App as presented in Fig. 9. Each time a patient uses this function it stores the voice sample of the cough for three times. The voice is captured via the smartphone's microphone and its voice signal readings stored into a single file.

## V. CONCLUSION

In this study, the novel coronavirus COVID-19 and the techniques for its detection have been discussed in details. Recent studies have also been investigated and explained in terms of their solutions from a range of radiology to the availability of modern technologies including medical detection Kit, CT image reading system and AI technologies. These solutions are either time consuming, costly or suffer from a lack of accuracy. However, this study proposed a novel framework to overcome such issues. Further, the proposed framework is based on using only smartphone sensors measurements and can be used by doctor or radiologists who have access to smartphones at anytime, anywhere. Further, the recommended or proposed framework could be run as an application on different smartphone-platforms, since it does not require any external or additional sensors providing higher accuracy.

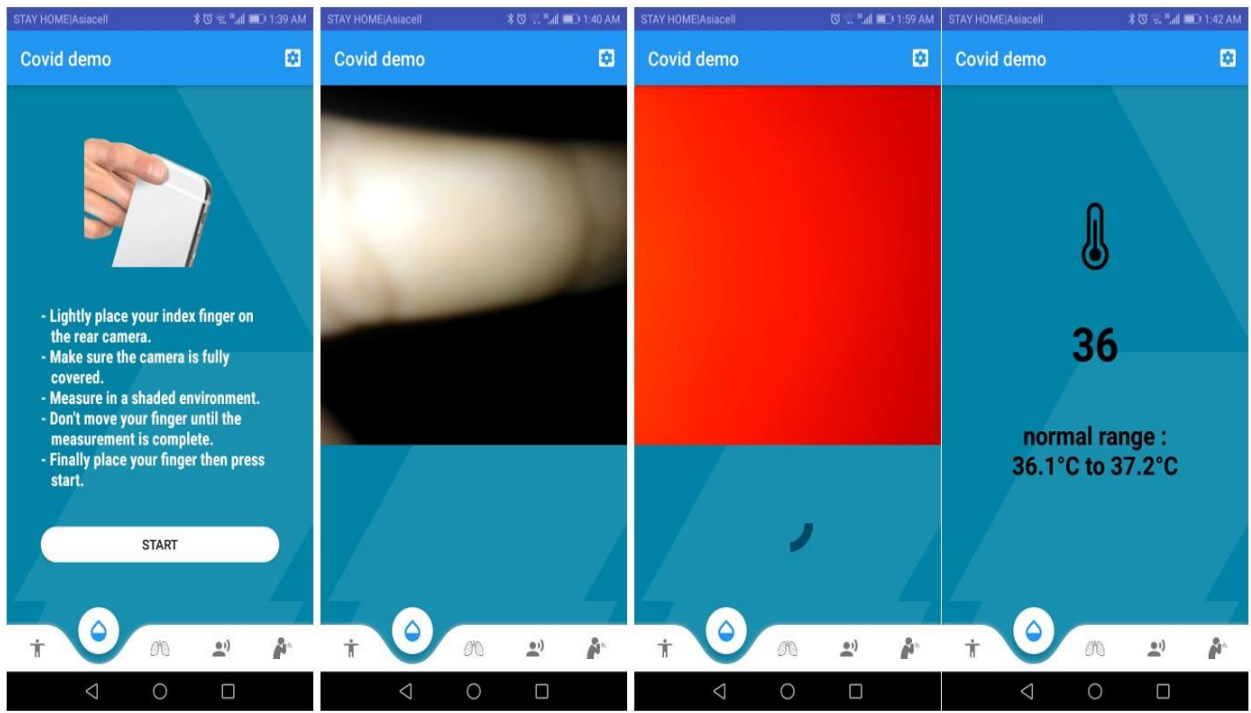

Fig. 6. Body temperature measurement.

The proposed framework includes four separate layers which are: input/reading sensors' measurements layer, sensors configuration layer, computing symptoms disease layer and predict the disease layer via using combined approaches. Also, the ML model in the final stage could be further improved by using a transfer learning method when the model has worked on the cloud. The framework is more reliable in comparison with the state-of-the-art; this is because, the framework relies on multi-readings from multiple sensors referring to the related symptoms of the disease. In future, we plan implement the designed framework on the smartphones.

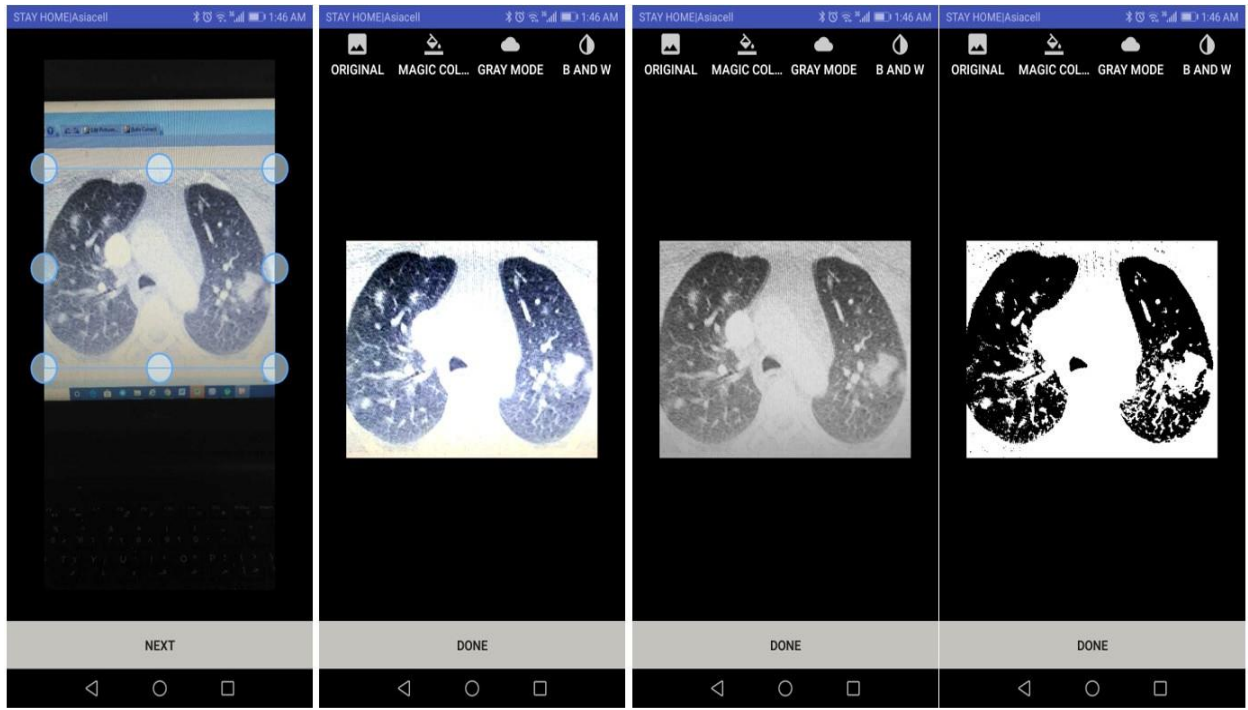

Fig. 7. Reading CT images.

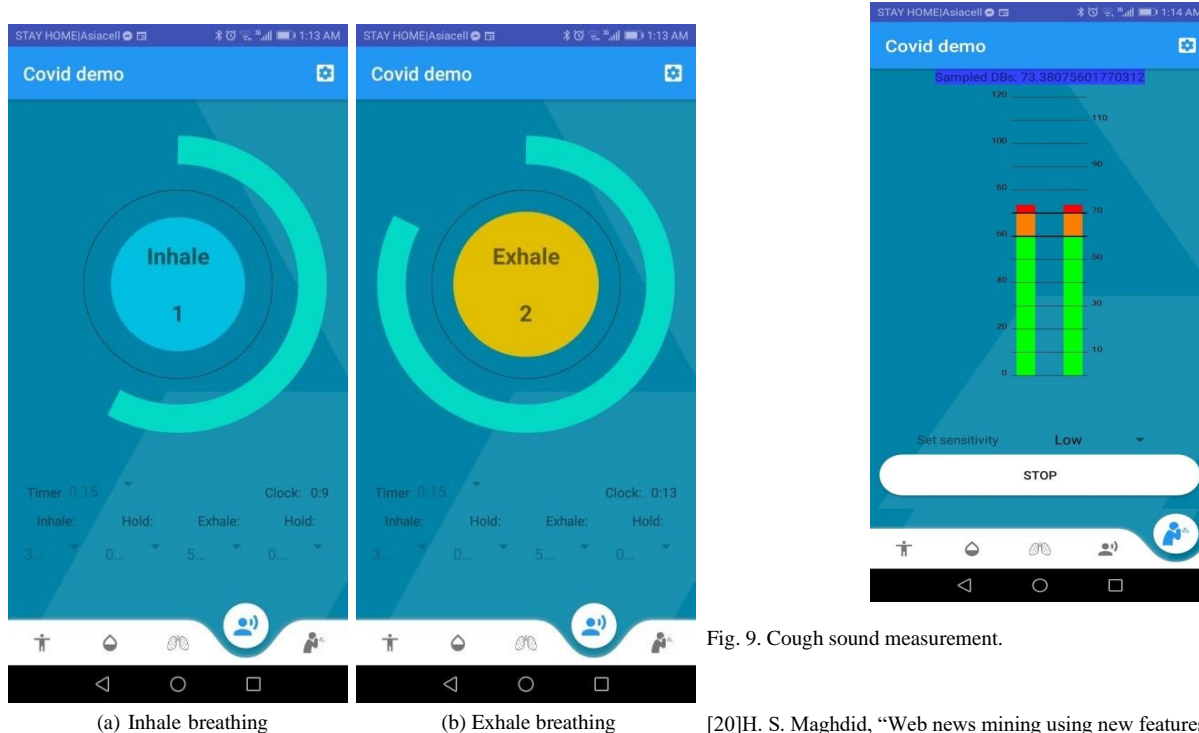

(a) Inhale breathing　　　　(b) Exhale breathing

Fig. 8. Breathing behavior analysis

Fig. 9. Cough sound measurement.